\documentclass[final,3p,times,twocolumn]{elsarticle}

\usepackage{amssymb}
\biboptions{numbers,sort&compress}
\usepackage[figuresright]{rotating}
\usepackage{amsmath}
\usepackage{float}
\usepackage{amsthm}
\usepackage{algorithm}
\usepackage{lastpage}
\usepackage{array}
\usepackage{enumitem}
 \usepackage{caption}
\biboptions{numbers,sort&compress}
\usepackage[thinlines]{easytable}
\newcommand*{\thead}[1]{\multicolumn{1}{>{\small}l}{\bfseries #1}}

\journal{Elsevier: Applied Energy}

\begin{document}

\begin{frontmatter}



\title{Short-Term Forecasting of CO\textsubscript{2} Emission Intensity in Power Grids by Machine Learning}


\author[DTU]{\small Kenneth Leerbeck\corref{cor1}}
\cortext[cor1]{{\scriptsize Corresponding Author \\
\textit{Email address:} kenle@dtu.dk \\
\textit{Postal Address:} Anker Engelunds vej 1, Building 101A, 2800 Kongens Lyngby, Denmark}}
\author[DTU]{Peder Bacher}
\author[DTU]{Rune Gr{\o}nborg Junker}
\author[DTU]{Goran Goranovi\'{c}}
\author[TMW]{Olivier Corradi}
\author[DTU]{Razgar Ebrahimy}
\author[DTU]{Anna~Tveit}
\author[DTU,NTNU]{Henrik Madsen}

\address[DTU]{Technical University of Denmark}
\address[TMW]{Tmrow IVS}
\address[NTNU]{Norwegian University of Science and Technology}

\begin{abstract}
A machine learning algorithm is developed to forecast the CO\textsubscript{2} emission intensities in electrical power grids in the Danish bidding zone DK2, distinguishing between average and marginal emissions. 
The analysis was done on data set comprised of a large number (473) of explanatory variables such as power production, demand, import, weather conditions etc. collected from selected neighboring zones. The number was reduced to less than 30 using both LASSO (a penalized linear regression analysis) and a forward feature selection algorithm. Three linear regression models that capture different aspects of the data (non-linearities and coupling of variables etc.) were created and combined into a final model using Softmax weighted average.
Cross-validation is performed for debiasing and autoregressive moving average model (ARIMA) implemented to correct the residuals, making the final model the variant with exogenous inputs (ARIMAX). The forecasts with the corresponding uncertainties are given for two time horizons, below and above six hours. Marginal emissions came up highly independent of any conditions in the DK2 zone, suggesting that the marginal generators are located in the neighbouring zones.

The developed methodology can be applied to any bidding zone in the European electricity network without requiring detailed knowledge about the zone.

\end{abstract}



    
    



\end{frontmatter}



\section{Introduction}
\label{Section:Introduction}
Consumption of electricity contributes heavily to the CO\textsubscript{2} emissions, \cite{IEA_emi}. The intensity with which it does so depends on the proportion of renewable sources (eg. solar, wind) vs. nonrenewable sources (eg. coal, gas, nuclear). The proportion, and hence the CO\textsubscript{2} emission, fluctuates with time based on power market mechanisms and weather conditions. Ideally, in the future, electricity users (the demand) would respond to the renewable power generation in attempt to lower the emissions. Proposed solutions include scheduling of storage (e.g. batteries, fuel cells, hydro reservoirs, thermal) and flexible demand (e.g. heat pumps, electric cars), \cite{IEA_sol}. This paper presents additional methodology aimed to fulfill the goal of lowering the emissions.

Forecasting of the power grid is essential for these solutions and exist in various forms already. For renewables, new forecasting methods are developed on a regular basis to increase revenue. Most recently in \cite{Zheng}, a multi-step ahead deterministic forecasting on wind power is done (many prior models only did one-step ahead), using complex multi-stage machine learning (kernel-based) algorithms for error correction. 
The deterministic model, however, cannot reliably account for the volatile wind speeds. Hence a probabilistic model (distribution forecast) of wind speeds based on robust machine learning algorithms was developed in \cite{Wang}. The probabilistic models quantify uncertainty of a forecast, crucial for risk management. On the other hand, multi-step forecasts are important in any scheduling application, such as market bids and flexibility. Examples of the combination of the two types of models are the multi-step ahead probabilistic solar power forecasts developed in \cite{Bacher} and  \cite{Delle} with horizons of 36 and 72 hours ahead respectively.



\begin{figure}[t]
\centering
  \includegraphics[width=\linewidth]{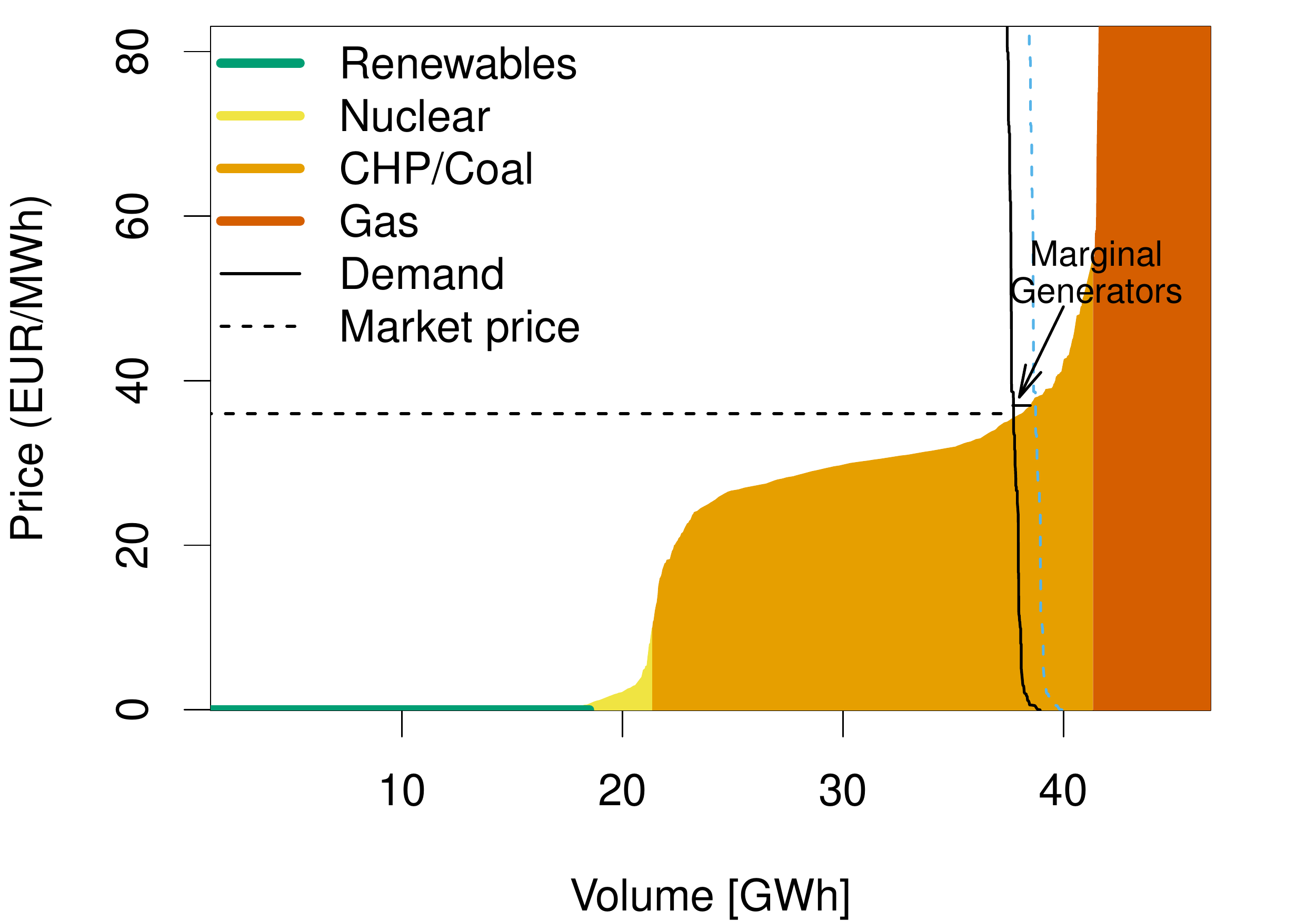}
  \caption{The merit order illustrated with a supply and demand curve example. The x-axis is the accumulated generators in the power system and the y-axis is their corresponding costs. The highest generator in the merit order is the one crossing with the demand curve - a coal generator in this example. The average emissions are a weighted average from all activated generators. The\textbf{ marginal generator} is the generator that will be activated by moving the demand line slightly to the right (dashed blue line). Data source: Nord Pool AS.}
  \label{fig:merit}
\end{figure}

The generator responding to small changes in demand is called the marginal generator and is not weather dependent (must generate on-demand) - hence, does not include wind turbines and PV-panels\footnote{\label{foot:mar}In modern electricity networks with high proportions of renewable sources, a weather-dependent generator \emph{can} become marginal. For example, during overproduction from wind turbines, demand cannot keep up and pushes down the electricity prices; the wind turbines can thus be shut-off to down-regulate the production. According to data supplied by Energinet Denmark, these situations occurred for 12.5\% of all the hours in 2017 and 2018 in Western Denmark (bidding zone DK1), where the marginal generator could be argued to be a wind turbine. This calls for a (future) refinement of the marginal CO\textsubscript{2} emission estimation.}. A good estimate of the marginal generator is achieved by using price signals, Figure~\ref{fig:merit}, where the marginal generator is, in this case, a coal fired generator.

Because price-based control can be both economically and environmentally beneficial, forecasting of the day-ahead spot prices has been proposed in recent papers \cite{Dogan,Zhang}, using different methodologies (e.g. ARIMA, Neural Networks, machine learning). There is a problem though with the spot prices, known as the merit order emission dilemma, illustrated for the German-Austrian power market \cite{Wolfgang}: The price for coal is low but the emissions are high. A price based-control, therefore, only leads to low emissions if there is surplus of renewable energy (more renewable energy than needed) - otherwise coal is favored.

Precise estimates and forecasts of the marginal CO\textsubscript{2} emissions are needed to correctly implement storage and flexible demand into the power grid. Discussing CO\textsubscript{2} emissions, two distinct concepts are used; average and marginal emission intensities $\left(\frac{\text{kgCO}_2\text{-eq}}{\text{MWh}}\right)$. Average emissions correspond to the overall, e.g. region-wide, electricity production including net imports. The marginal reflects the emissions of the marginal generator. The concepts are compared in \cite{Anika} and the importance of distinguishing between the the two is highlighted due to their very opposing patterns. 

Both concepts have been estimated in prior studies \cite{Voors,Voors2,Marnay,Beetle,Hawkes,Rekkas,Hadland,CO2_signal_heating,Corradi,Anika}. Early studies of the marginal emissions,  \cite{Voors,Voors2,Marnay,Beetle}, all estimate the highest generator in the merit order of the power generation system. However, this is rarely the only generator responding to a change in demand, as addressed in \cite{Hawkes}, where these early approaches are discussed and a new empirical approach is presented; By estimating the contribution of all power generators to a specific change in system demand, using linear regression on historical data including outputs from major power producers in the UK to estimate the average response from each generation technology class to changes in demand. The power plants were disaggregated to investigate the impact of plant turnovers (switching old power plants for new ones).
Traditionally treated as dummy variables, the imports have been treated explicitly in a recent study that focused on the average emissions in the Nordic European countries, \cite{CO2_signal_heating}. The study showed the interplay between the imports and the average emissions and how both vary from one bidding zone to another. Incorporating the imports in the marginal emissions, the company Tmrow IVS has developed a new empirical approach using machine learning on historical data that follows the chain of imports (the so-called flow tracing, originally introduced in \cite{Bialek,Kirschen}) to assess the impact of a specific generator or load on the power system \cite{Corradi}. This is a large scale solution using data from the majority of bidding zones around the world. 

The just mentioned studies provide methodologies for marginal CO\textsubscript{2} emission \emph{estimates}. Also, the \emph{long-term} (e.g. annual) forecasting of CO\textsubscript{2} emissions are widely conducted for promoting green energy, e.g. \cite{Azim}. However, the more accurate short-term emission forecasting methodologies, currently unavailable in the literature, are needed to implement the flexible demand. 
In this study, short-term (24h ahead) forecasts with uncertainty margins (95$\%$ prediction intervals) of both the average and marginal CO\textsubscript{2} emission intensities from the power generation in bidding zone DK2 (Sealand region, Denmark) are developed. These enable flexible consumers (electric cars, heat pumps, etc.) to schedule for optimal electricity usage i.e minimal CO\textsubscript{2} emissions, be it for regulatory or branding purposes. The methodology can be applied to other bidding zones in the European electricity network without requiring detailed knowledge of response and explanatory variables.

The forecasting in this study models the given response variables (average and marginal CO\textsubscript{2} emission intensities) in terms of so-called explanatory variables - e.g. power production, demand, import, weather conditions, etc. - that are represented by generalized basis functions (splines). Examples of the explanatory variables with respect to the response variables are shown in Sec.~\ref{Section:Explore}. The data is divided into two sets: one set available for $\leq 6$ hours and another available for horizons $> 6$ hours. The machine learning techniques that include trend extraction (seasonality, nonlinearity, interaction terms), feature selection (LASSO, forward feature selection), residual correction (ARIMA) and cross-validation are elaborated in Sec.~\ref{Section:Model_tech} and \ref{sec:ML2}, where also three different models are built and combined with a Softmax weighted average into the overall forecasting model. The most significant variables and forecasting results are highlighted in Sec.~\ref{section:Results}, and concluding remarks are found in Sec.~\ref{Section:Conclusion}. A list of all the variables can be reviewed in \ref{Appendix:Variables} where it is indicated which variables are being used for which set of data.

\section{Data analysis: examples}
\label{Section:Explore}

The CO\textsubscript{2} emissions in Denmark are interesting when scheduling flexible consumers due to large amounts of wind power production - 48\% of the total electricity production in 2017. The country also has a variety of good trading options with its neighboring countries - e.g. Germany (DE), Sweden (SW) and Norway (NO). Indirectly, influence can also come from countries further away, but the scope of this study is limited to selected bidding zones DE, DK1, DK2, NO2, SW3 and SW4. The focus area, DK2, has direct transmission cables to DK1, SW4 and DE.

\begin{figure}[t]
\centering
  \includegraphics[width=\linewidth]{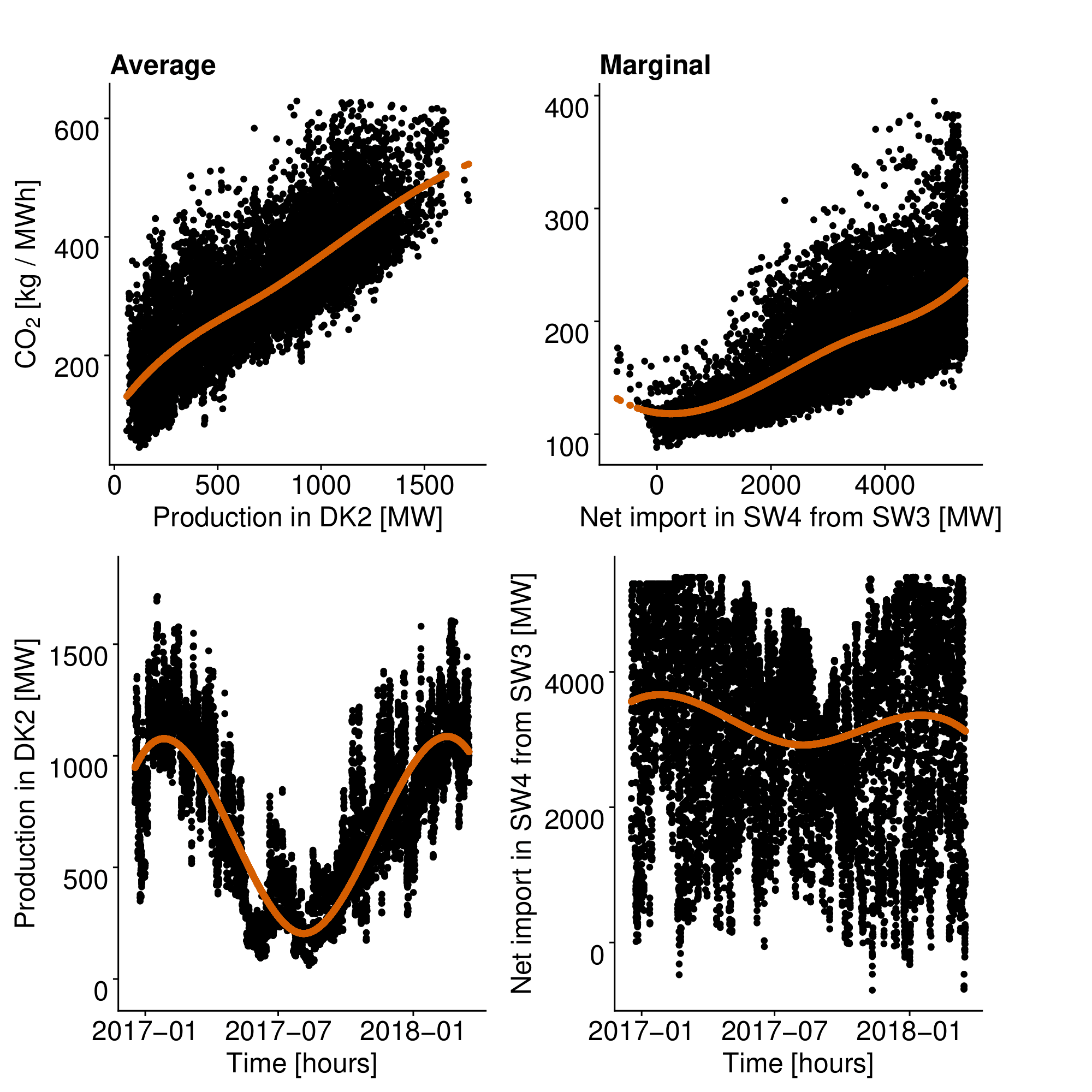}
  \caption{Top left: The average CO\textsubscript{2} emission intensity vs the power production in DK2. Top right: The marginal CO\textsubscript{2} emission intensity vs the net import in SW3 from SW4. Bottom: The power production in DK2 and the net import in SW3 from SW4 vs time (hourly resolution).}
  \label{fig:pair}
\end{figure}

Linear relationships between the response and explanatory variables are detected first. The power production in DK2 and net import in SW4 from SW3 show the highest correlations to the response variables as illustrated in Figure \ref{fig:pair}. The average emissions are highly correlated to the power production in DK2 because it is a response to the demand and this will mainly activate coal and gas-fueled generators. 
Net import in SW4 from SW3 show the highest correlation to the marginal emissions and is an indirect influence - this shows that the marginal generator is often located in SW3 where all Sweden's nuclear power is produced \cite{Kraftnat}. This is because nuclear power is cheaper than the local options in DK2 e.g. coal and gas. Figure \ref{fig:pair} also shows a yearly seasonality in the local power production varying about 1,000 MW - lowest in the summer. The import show the same yearly seasonality but less significant because nuclear power serves the baseload.

\begin{figure}[t]
\centering
  \includegraphics[width=\linewidth]{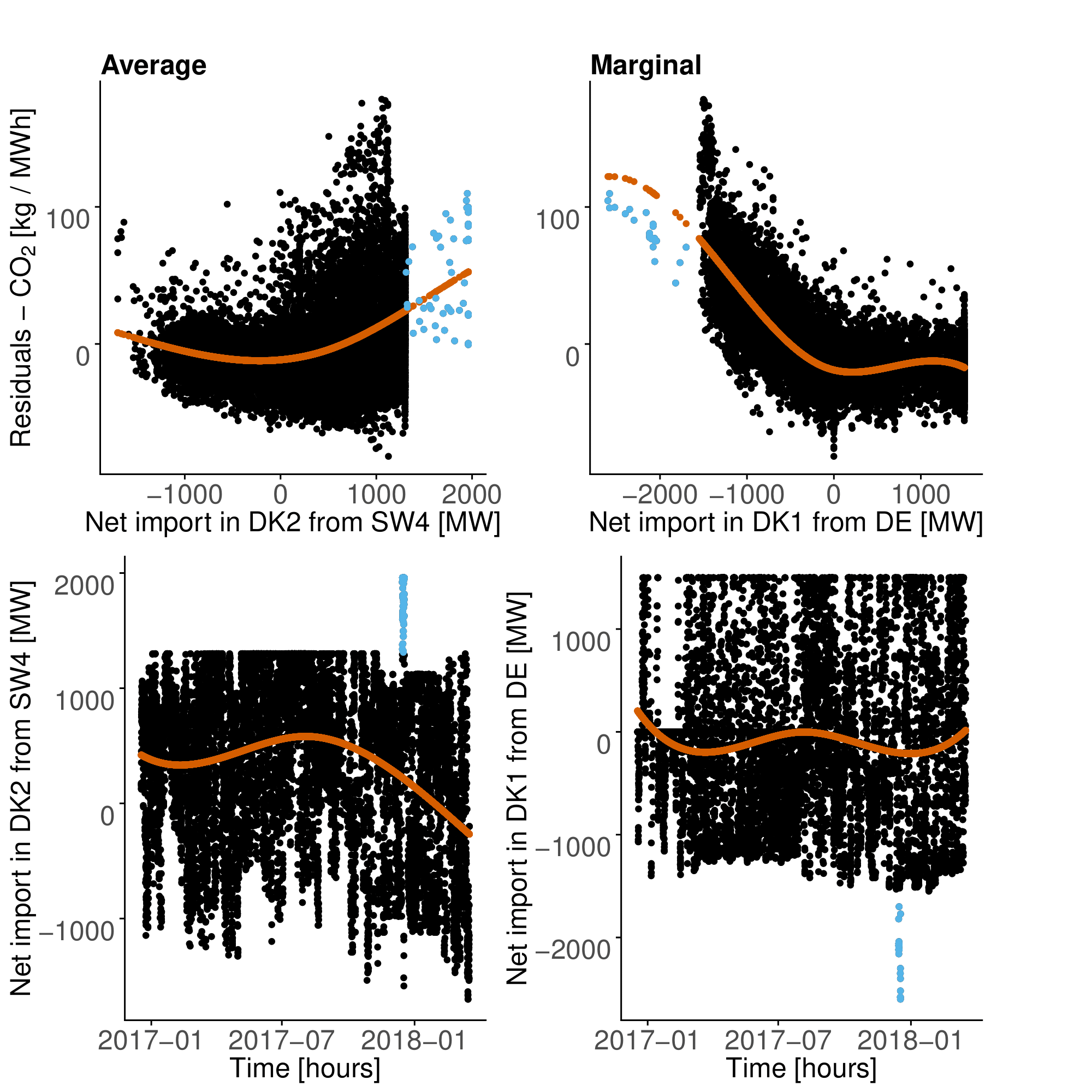}
  \caption{Top left: The residuals from the linear fit of the domestic production (DK2) onto the average CO\textsubscript{2} emission intensity vs the net import in DK2 from SW4. Top right: The residuals from the linear fit of the import in SW4 from SW3 onto the marginal CO\textsubscript{2} emission intensity vs the net import in DK1 from DE. Bottom: The net import in DK2 from SW4 and the net import in DK1 from DE vs time (hourly resolution).
  Note the outliers (light blue); over three days, the import seemingly overloaded both the transmission cables to Germany and Sweden. For this study, these data points are modified to the maximum capacity (1,300 and 1,550 MW in these cases)}
  \label{fig:pair_res}
\end{figure}

Next, a linear regression model can be fitted onto the average and marginal emissions using the discussed production and import respectively to reveal other important variables. The residuals from these models will represent the average and marginal emissions that are independent of these variables.
This reveals the non-linear relationships shown in Figure \ref{fig:pair_res}. 

Interestingly, the average emissions are the highest when the net import from SW4 is zero, indicating trades usually happen when cheap non-polluting electricity is generated - e.g. nuclear and renewables - and the corresponding CO\textsubscript{2} emissions are proportionally lowered. This extra power, produced e.g. in Sweden, can then be imported for use and serves as an indicator of the lower CO\textsubscript{2} emissions. The marginal emissions are highly depending on the net import in DK1 from DE but only when DK1 exports to DE. The higher the export the more domestic generators must serve as marginals. 
The net import in DK2 from SW4 shows a yearly seasonality being the highest in the summer - recall the domestic production being the lowest here - since demand is generally low the proportion of nuclear power is large and Denmark can, therefore, import it rather than produce electricity locally from gas or coal.

\section{Regression models and basis functions}
\label{Section:Model_tech}



\subsection{Linear Regression Models (lm)} \label{Appendix:lm}
The CO\textsubscript{2} emissions are modeled using a multivariate linear regression model
\begin{align}\label{eq:linear}
    \mathbf{Y} &= \mathbf{X}\mathbf{\beta} + \epsilon, \\
    \text{for} \; \epsilon &\sim N(0,\sigma^2 I), \nonumber
\end{align}
where $\mathbf{X}$ is the input matrix (explanatory variables), $\mathbf{Y}$ is the output vector (response variables: CO\textsubscript{2} emissions) and $\mathbf{\beta}$ is a vector of regression coefficients to be found. $\epsilon$ represents the normally distributed errors in the model. 

The least square regression is performed to minimize
$$ S(\mathbf{\beta}) =  ||\mathbf{Y} - \mathbf{X}\mathbf{\beta}||^2 $$
and obtain the ordinary least-squares solution
\begin{equation}
\mathbf{\beta} = (\mathbf{X}^T \mathbf{X})^{-1} \mathbf{X}^T \mathbf{Y}.\label{eq:coeff}
\end{equation}

Once $\mathbf{\beta}$ is obtained on training data, the response variables $\hat y$ can be forecasted on new input data (the test data)

\def\horizontaldistance{\kern3.3pt}

\begin{align}
     \hat{\mathbf{y}}_{t+h} =  
\begin{bmatrix}
  \kern0.15pt \mathbf{x}^\text{F}_{t+h} \horizontaldistance MA\left(\mathbf{x}^\text{RT}_{t+h}\right)_{24} \horizontaldistance MA\left(\mathbf{x}^\text{RT}_{t+h}\right)_{48} \horizontaldistance lag(\mathbf{y})_{t+h} \kern0.15pt  
\end{bmatrix}\mathbf{\beta} \equiv \mathbf{z}^*_{t+h}~\mathbf{\beta},
\label{eq:forecast}
 \end{align}
where F and RT refer to Forecast and Real-Time, and MA to Moving Average. The terms in the brackets (the matrix elements) are provided by Tmrow: $\mathbf{x}^\text{F}_{t+h}$ are the input variables forecasted $h$-hours ahead ($h\leq 6$ hours for all explanatory variables except weather data which is available for $h\geq 6$ hours). The moving average is constructed to translate the real-time input variables into the forecasting format
\begin{equation}
    MA\left(\mathbf{x}^\text{RT}_{t+h}\right)_n =  \frac{1}{n} \sum_{i = 0}^{n-1} \mathbf{x}^\text{RT}_{t - i},
\end{equation}
where $n$ is the length of the averaging period - either 24 or 48 hours. Finally, $lag(\mathbf{y})_{t+h}$ in Eq.~\ref{eq:forecast} is just $\mathbf{y}_t$, the last available observation at time $t$. The idea is to include all available information to obtain the forecast. This is an Auto-Regressive (regression on lagged values of $\mathbf{y}$) model with eXogeneous inputs (regression on functions of $\mathbf{x}$) or ARX for short.


Both $\mathbf{X}$ and $\mathbf{Y}$ are given as time series. $\mathbf{X}$ consists of 418 variables - for $h\leq 6$ hours - (related both to weather and power system from six biding zones; 66 listed in \ref{Appendix:Variables}) and $\mathbf{Y}$ of 10,897 observations. The forecasted input variables at $t+h$ constitute a new input matrix $\mathbf{Z}$ that will be expanded with more columns (next three subsections).
\begin{figure}[t]
\centering
  \includegraphics[width=\linewidth]{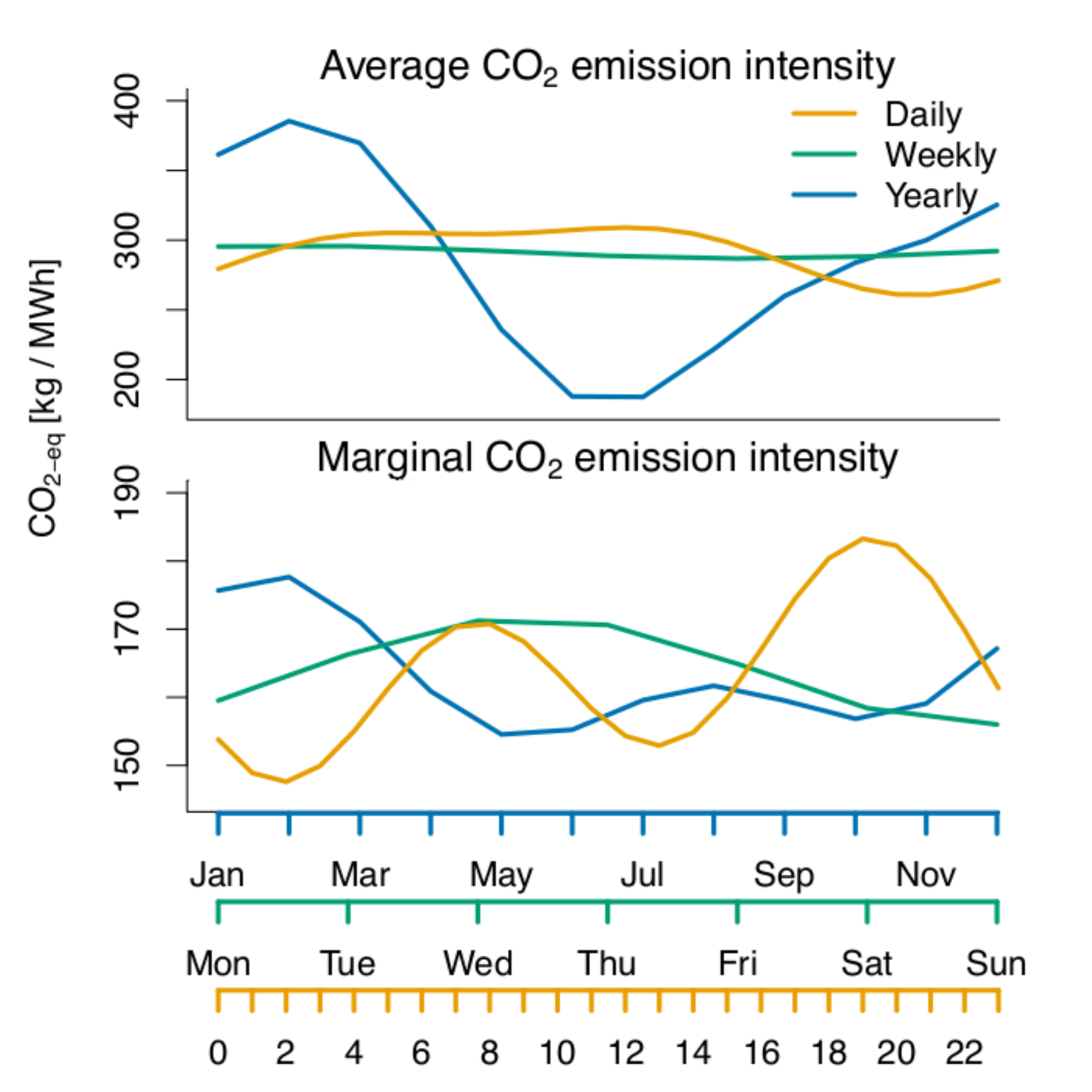}
  \caption{Fourier Series showing the daily, weekly and yearly patterns for the average and marginal CO\textsubscript{2} emission intensities.}
  \label{fig:daily}
\end{figure}

\begin{figure}[t]
\centering
  \includegraphics[width=\linewidth]{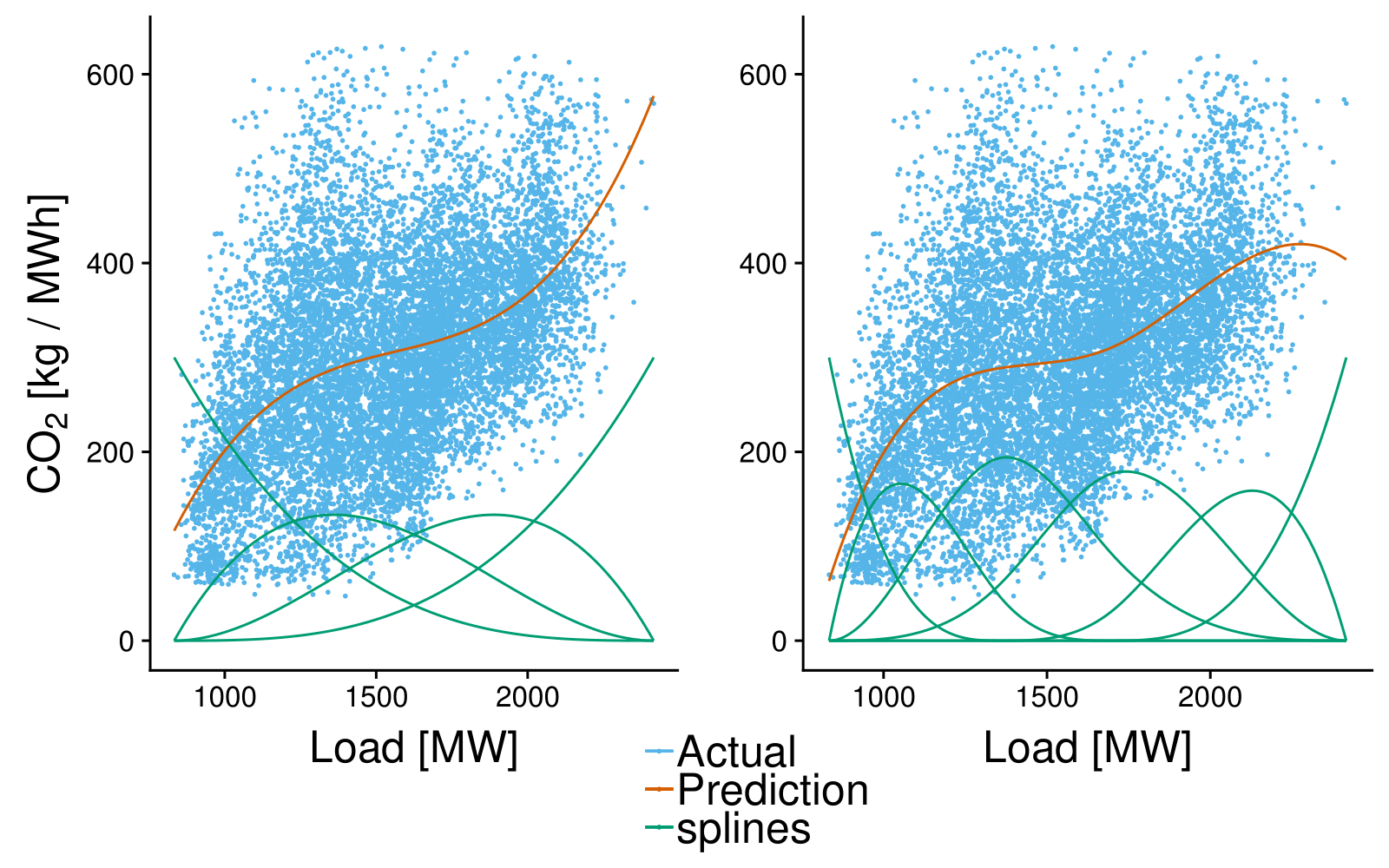}
  \caption{Estimation of the average CO\textsubscript{2} emission intensity (orange) based on four and six splines, respectively. The splines are scaled for illustration.}
  \label{fig:splines}
\end{figure}

\subsection{Periodic variations: Fourier Series}\label{section:periodic}
Periodic variations (seasonality) of the average and marginal CO\textsubscript{2} emission intensities are investigated using Fourier Series defined as
\begin{align}
    &FS(n,period)_t = \hat{\mathbf{y}}_{FS(n,period),t} =\\  &A_0 + \sum_{i = 1}^{n_{\text{series}}} A_i \cdot \sin \left(i \frac{2\pi t}{\text{period}}\right) + B_i \cdot \cos \left(i \frac{2\pi t}{\text{period}}\right), \nonumber 
\end{align}
where $\mathbf{y}$ is the response variable, $t$ is the time, $A$ and $B$ are linear regression coefficient matrices, $n$ is the order of the Fourier Series and 'period' is the length of the seasonality period. $n$ is adjusted to find the best fit. In Figure \ref{fig:daily} the daily, weekly, and yearly patterns are estimated using data ranging from January 2015 to January 2018 with $n = [2, 1, 2] $ respectively.

It is observed that average and marginal emissions follow completely different patterns. The yearly pattern shows the largest variations for the average emissions, as already discussed in Section \ref{Section:Explore}. 
The daily average emission varies very little in comparison to the yearly emissions. On the other hand, the daily marginal emissions are the highest when the average emissions are the lowest, illustrating the importance of using the correct emission measure. The weekly pattern is only of importance to the marginal emissions and is lowest on the weekends. The seasonality components are added to the input matrix $\mathbf{Z}$ of Eq. \ref{eq:forecast} such that
\begin{align}
    \mathbf{z}_t =  \begin{bmatrix}
\mathbf{z}^*_t & exp^{\mathbf{z}^*_t} & FS(2,24)_t & FS(1,7)_t & FS(2,12)_t
\end{bmatrix},
\end{align}
Note that an exponential term of $\mathbf{Z}$ is added, too, which makes $Z$ a (10,897x951) matrix (both $\mathbf{Z^*}$ and $\exp{(\mathbf{Z^*})}$ have 474 columns after the clean-up of non-available points and constant variables).

\subsection{Non-linearities: Splines}\label{section:Non-linear}
Non-linearities are captured by using splines, the local polynomials between specified points called knots \cite{Springer}.
Splines are implemented in \textsf{R} with the built-in functions \texttt{bs()} (base splines) and \texttt{ns()} (natural splines). The former are basis functions and increasing their number in the expansion of a function improves the fitting procedure at the risk of overfitting. In this study, four base splines with knots located at the quantiles (default settings in \textsf{R}), are used to avoid overfitting
\begin{equation} \label{eq:splines}
 \mathit{bs}(\mathbf{z}_t) =
 \begin{bmatrix} \mathit{bs}_0(\mathbf{z}_t) & \mathit{bs}_1(\mathbf{z}_t) & \mathit{bs}_2(\mathbf{z}_t) & \mathit{bs}_{3}(\mathbf{z}_t)
 \end{bmatrix}^{\top}.
 \end{equation}
The number is justified by the assumption that the relationships between the explanatory and the response variables are stationary. The least-square coefficients associated with the base splines are labeled by the vector $\mathbf{\beta^{\emph{bs}}}$. To refine the fitting, natural splines $\mathit{ns}(\mathbf{z}_t)$ are also used and represented analogously to Eq. \ref{eq:splines}.

Figure \ref{fig:splines} shows the estimated mean (orange) of the average CO\textsubscript{2} emissions in DK2 calculated on the basis of the demand, based on four and six splines (green).

\subsection{Interaction terms} \label{sec:Interactions}
The interactions \cite{Aike} between the explanatory variables is modeled as 
\begin{equation} \label{eq:interaction}
 \mathit{IA}(\mathbf{z}_{i,t}, \mathbf{z}_{j,t}) =
 \begin{bmatrix} 
 \mathbf{z}_{i,t} & \mathbf{z}_{j,t} &  \mathbf{z}_{i,t} \cdot \mathbf{z}_{j,t}
 \end{bmatrix}^{\top},
\end{equation}
where the product $\mathbf{z}_i \cdot \mathbf{z}_j$ denotes the mutual interaction of the $(i,j)$ pair. The coefficients are denoted as vector $\mathbf{\beta}^{\emph{IA}}$.

Interactions were also represented by splines to refine the non-linearity, i.e.
\begin{equation}
    bs(\mathit{IA}(z_{i,t},z_{j,t}))  = 
 \begin{bmatrix}
  \mathit{bs}_0(\mathbf{z}_{i,t}) & \mathit{bs}_0(\mathbf{z}_{j,t}) & \mathit{bs}_0(\mathbf{z}_{i,t}\cdot \mathbf{z}_{j,t}) \\ 
  \mathit{bs}_1(\mathbf{z}_{i,t}) & \mathit{bs}_1(\mathbf{z}_{j,t}) & \mathit{bs}_1(\mathbf{z}_{i,t}\cdot \mathbf{z}_{j,t}) \\  
  \mathit{bs}_2(\mathbf{z}_{i,t}) & \mathit{bs}_2(\mathbf{z}_{j,t}) & \mathit{bs}_2(\mathbf{z}_{i,t}\cdot \mathbf{z}_{j,t})\\
  \mathit{bs}_3(\mathbf{z}_{i,t}) & \mathit{bs}_3(\mathbf{z}_{j,t}) & \mathit{bs}_3(\mathbf{z}_{i,t}\cdot \mathbf{z}_{j,t})
 \end{bmatrix}^{\top},  \label{eq:interaction_splines}
\end{equation}
with the corresponding coefficients the matrix $\mathbf{\beta}^{\emph{bs(IA)}}$.
 
Note that explanatory variables generally change in time; for example, the production in DK2 has a clear daily pattern and its corresponding linear regression coefficient will vary accordingly. This can be expressed as interactions with the time variables (hour, week, month). A separate matrix $\mathbf{\tau}$ is thus defined to group the periodic as well as nonlinear character of the time variables (\ref{Appendix:tau}).

\section{Statistical selection and refinement of models}\label{sec:ML2}

\subsection{Cross validation strategy} \label{section:CV}
Throughout the study rolling forward cross-validation is used, where data is divided into eight sets each consisting of training, validation and testing data \cite{Hu}. There are eight rounds of cross-validation, where the training data is increased by one set in each round (the validation set of a previous round becomes part of the training data in the next one), and the validation and testing sets are always new independent data sets. 

The cross-validation is done by averaging the Root Mean Squared Error (RMSE) over the eight validation sets. The final comparison of models is made on the test sets. The data range is 15 months, from 19th December 2016 to 18th March 2018. This is a smaller period from the one mentioned in seasonality, Sec. \ref{section:periodic}. The reason is that many of the explanatory variables have limited historical data and hence cannot be cross-validated further back.\\

\subsection{Feature selection techniques}\label{section:feat_select} 
\begin{figure}[t]
\centering
  \includegraphics[width=\linewidth]{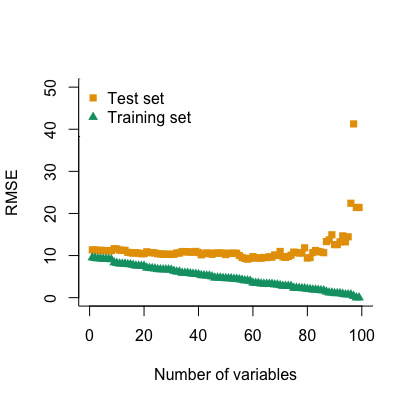}
  \caption{RMSE vs number of variables. Generated using 200 observations from 100 independently simulated explanatory variables, all with a certain degree of correlation to a simulated response variable.}
  \label{fig:overfitting}
\end{figure}

Feature selection is necessary to remove co-linearity and overfitting in linear regression. The co-linearity happens when two or more explanatory variables are linearly dependent or highly correlated. When this happens the condition number of the matrix $\mathbf X$ is lowered, making the determinant of it close to zero, and thus inverting it results in large numerical errors. The overfitting is related to a large number of model parameters used to fit training data, which then causes poor model performance on test data, as seen from the yellow points in Figure \ref{fig:overfitting}. In the extreme, high order polynomials are found to perfectly fit scattered data that in reality follow a simple, say, linear trend.

Two methods are used for feature selection. In the first, Least Absolute Shrinkage and Selection Operator (LASSO) algorithm, a penalty term is added to the objective function $S$ that is to be minimized
\begin{equation}
    S(\beta) =  ||\mathbf{y} - \mathbf{X}\mathbf{\beta}||^2 + \lambda|| \mathbf{\beta}||_{1},\label{eq:object_funct_lasso}
\end{equation}
where $\lambda$ is the penalty parameter and the subscript $1$ indicates the L1 norm; the larger the L1 norm of the coefficient $\mathbf{\beta}$, the larger the penalty \cite{Tibshirani}. This reduces parameter estimates to zero, hence its name "Shrinkage".


The penalty term $\lambda$ is tuned using the average RMSE from the validation sets defined in the above eight-fold cross-validation strategy. The higher the value of $\lambda$ the more the $\beta$'s shrink towards zero and therefore fewer variables will be selected. $\lambda$ is slowly decreased from an initial high value until the optimal value is reached when model performance stops improving. 
If the performance starts to decrease, the model is over-fitted.

Before applying LASSO, highly correlated variables are removed manually to reduce the computation time of the LASSO regression.

The second method is the forward feature selection algorithm in which new variables are added to the best models and tested for improvement, \ref{Appendix:FS}. These methods are used for this study in the following order
\begin{enumerate}
    \item  Highly correlated variables ($\rho >0.99$) are removed.
    \item  LASSO regression is applied.
    \item  The forward selection algorithm is applied.
    \item  Step two and three repeated with updated variables until convergence (number of variables does not decrease anymore). 
\end{enumerate}

\begin{figure}[t]
\centering
  \includegraphics[width=\linewidth]{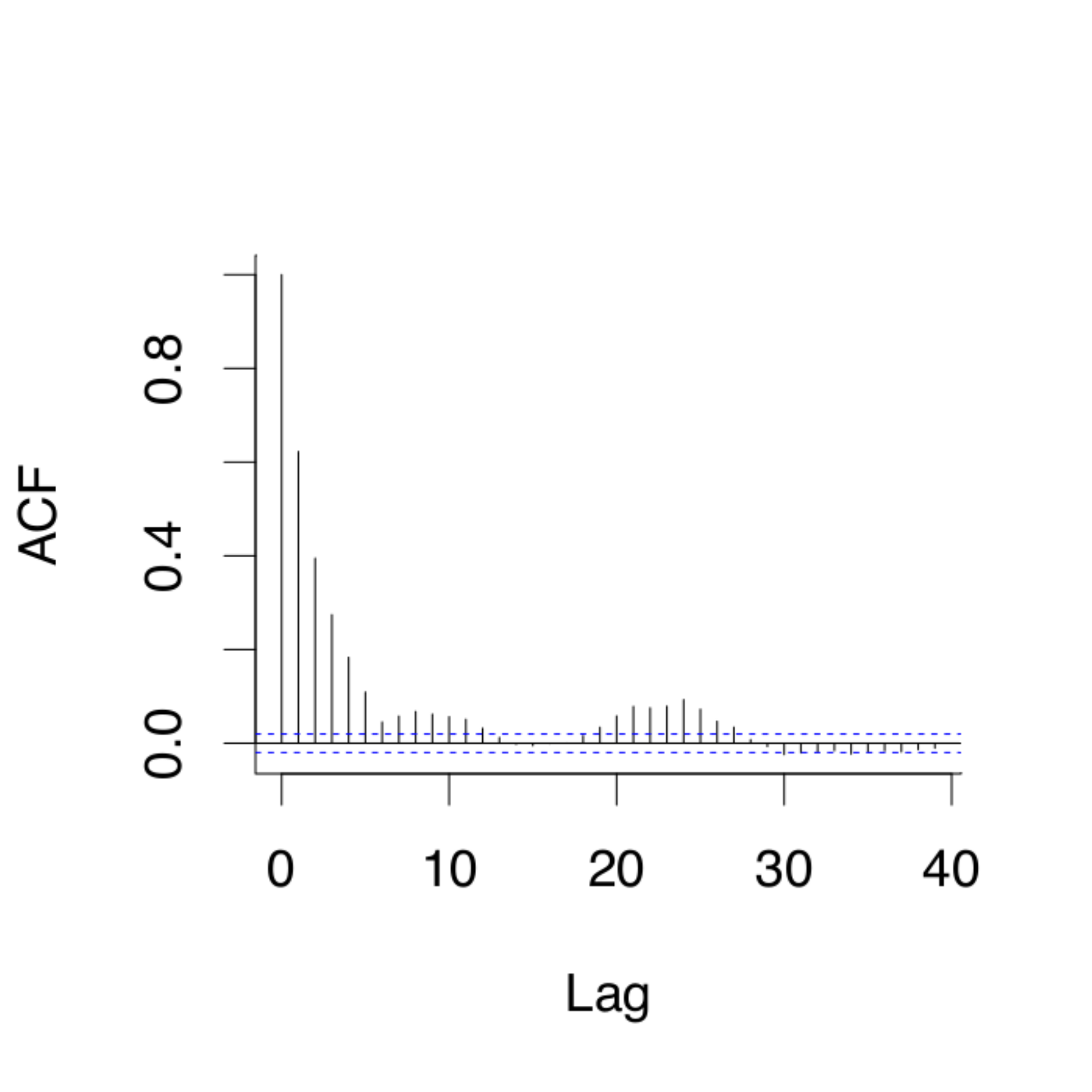}
  \caption{Auto Correlation Function of the residuals from the combined model - $\mathbf{y}$ is the average CO\textsubscript{2} emission intensity and $h = 6$ hours - on the training set from CV set 8.}
  \label{fig:acf1}
\end{figure}

\subsection{Residual correction}\label{section:corrections}
When the model does not predict well the test data, residuals are non-random and become auto-correlated (correlated to lagged versions of itself). Thus corrections to the model need to be made to account for the correlation. 

Residual auto-correlations are checked with the ACF (Auto Correlation Function) that reveals any linear dependencies in the residuals. In Figure \ref{fig:acf1}, residuals of the compound model (see later, Sec. \ref{section:weights}) of the average CO\textsubscript{2} emission intensity on six-hour horizon is shown. There are high correlations up until the lag of six hours, and also a smaller correlation around 24 hours due to seasonality.

The residual was modeled independently with Auto Regressive Moving Average (ARIMA) model \cite{TSA} 
\begin{equation}
    \mathbf{Y}_{t} = \epsilon_{t} + \psi_1\epsilon_{t-1} + \cdots + \psi_q \epsilon_{t-q} - \phi_1 \mathbf{Y}_{t-1} - \cdots - \phi_p \mathbf{Y}_{t-p} ,
\end{equation}
containing $p$ lagged values (AR part) and $q$ errors of previous observation of the moving average (MA). Considering an AR model the prediction errors are obtained as
\begin{equation} \label{eq:arima_pred}
    \mathbf{Y}_{t+h}-\mathbf{\hat{Y}}_{t+h|t} = \epsilon_{t+h} + \psi_1\epsilon_{t-1+h} + \cdots + \psi_{h-1} \epsilon_{t+1} .
\end{equation}

The models of this type are denoted ARIMA\textsubscript{p,d,q} process, where parameters $p$, $d$, $q$ refer to the AR, I and MA part, respectively - I is an integrating term used to make the data stationary (mean and variance are constant over time). In this study an extension is used, $ARIMA_{(p,d,q)(P,D,Q)_{M}}$, where $P$,$D$ and $Q$ are the seasonal parameters and $M$ is the length of the season. The parameters are fitted using information from the ACF and using the built-in function \texttt{auto.arima()} in \textsf{R}, which automatically selects the model with the best fit. In this case, it is found to be seasonal $ARIMA_{(3,0,0)(0,1,2)_{24}}$ and $ARIMA_{(5,1,0)(2,0,0)_{24}}$ models that removes all significant correlations for the average and marginal emission models with $h = 6$ hours.

The prediction of the residual model is added to Eq. \ref{eq:forecast} to obtain the final prediction
\begin{align} \label{eq:ARIMA}
         \mathbf{y}_{t+h|t} = \mathbf{\hat{y}}_{t+h} + ARIMA(\mathbf{y}_{1:t} - \mathbf{\hat{y}}_{1:t},h) + \epsilon_{t+h|t}
\end{align}
where $\mathbf{y}_{t+h|t}$ and $\epsilon_{t+h|t}$ are the response variable and error at time $t+h$ given (residual) information at $t$, and $\mathbf{\hat{y}}_{t+h}$ is the prediction from the linear model. 
$ARIMA(\mathbf{y}_{1:t} - \mathbf{\hat{y}}_{1:t},h)$ is the ARIMA predicted error in line with Equation \ref{eq:arima_pred}, however with an extended version.
Note that once obtained, the ARIMA model is used for predicting the residuals at the horizon $h$. Finally, the uncertainty of the model is evaluated by applying the 95\% prediction interval. This is in accordance to the equations defined in \cite{TSA}, and is applied easily in \textsf{R} through built-in options in \texttt{lm()} and \texttt{arima()}.

The described residual correction can improve both the forecast for the specific horizon and the forecast for the lower horizons at the same time. Besides this, it also gives more consistent results since the cross-validation sets do not stand out (the variance of the errors becomes smaller).

\subsection{Base models}
The formalism of Sections \ref{Section:Model_tech}, \ref{section:CV} and \ref{section:feat_select} is used to assemble different models of increasing complexity, listed below. Starting point is the model
\begin{itemize}
    \item \textbf{M0}
\begin{align}
     \mathbf{y}_{t+h} =\beta_0 &+ \sum_{i = 1}^{477} \beta_{i} (\mathbf{z}_{i,t+h}) + \epsilon_{t+h}, \nonumber
\end{align}
\end{itemize}

where the 477 refer to the columns in $\mathbf{Z}$.

\begin{itemize}
    \item \textbf{M1}
\begin{align}
     \mathbf{y}_{t+h} =\beta_0 &+ \sum_{i = 1}^{951} \beta_{i} (\mathbf{z}_{i,t+h}) + \sum_{i = 1}^{15} \beta^\tau_{i} (\tau_{i,t+h})\nonumber\\
     &+ \sum_{i = 1}^{477}  \beta^{bs}_{i} \cdot \mathit{bs}(\mathbf{z}_{i,t+h})
    + \sum_{i = 1}^{477} \beta^{ns}_{i}\cdot \mathit{ns}(\mathbf{z}_{i,t+h})  + \epsilon_{t+h} \nonumber
\end{align}
\end{itemize}
includes time variables $\tau$ and non-linearities (splines).
\begin{itemize}
    \item \textbf{M2}
\begin{align}
     \mathbf{y}_{t+h} = \beta_0 &+  \sum_{i=1}^{50} \sum_{j=1  \wedge j \neq i}^{50} \beta^{IA}_{i,j,k} \cdot \mathit{IA}(\mathbf{z}_{i,t+h},\mathbf{z}_{j,t+h})_k\nonumber \\
   &+ \sum_{i=1}^{50} \sum_{k=1}^{15} \beta^{IA}_{i,k,l}\cdot \mathit{IA}(\mathbf{z}_{i,t+h},\tau_{k,t+h})_l + \epsilon_{t+h}\nonumber
\end{align}

\end{itemize}
is based on the reduced number of features (maximum 50, obtained from the LASSO regression of the Model 0 and ranked based on the size of their linear regression coefficients), and their interactions defined by the vector $\mathit{IA}$, Equation \ref{eq:interaction}. 
\begin{itemize}
\item \textbf{M3}
\begin{align}
    \mathbf{y}_{t+h} = \beta_0 &+  \sum_{i=1}^{50} \sum_{j=1  \wedge j \neq i}^{50} \beta^{bs(IA)}_{i,j,l,m} \cdot bs\left(\mathit{IA}(\mathbf{z}_{i,t+h},\mathbf{z}_{j,t+h})\right)_{l,m} \nonumber\\
   &+ \sum_{i=1}^{50} \sum_{k=1}^{21} \beta^{bs(IA)}_{i,k,l,m}\cdot bs\left(\mathit{IA}(\mathbf{z}_{i,t+h},\tau_{k,t+h})\right)_{l,m} + \epsilon_{t+h}, \nonumber
\end{align}
\end{itemize}
is also based on the reduced features, but with the interactions defined via matrix $\mathit{bs(IA)}$ of Equation \ref{eq:interaction_splines}. 

The feature selection procedure defined in Sec. \ref{section:feat_select} is applied to \textbf{M1}, \textbf{M2} and \textbf{M3} and reduces the number of variables to 10-30 depending on the horizon.


\subsection{Weighted average model}\label{section:weights}
The final model was the weighted average of the above Models 1-3. The weights were based on the performance of models on eight validation sets. The Softmax function was used $$ w_i = \left(\frac{\exp{x_i}}{\sum_{j = 1}^{n} \exp{x_j}}\right),$$ 
where $w$ is the weight vector, $x$ is a vector representing the average $\mathit{RMSE}$ scores of the included models on the validation sets and $n$ is the number of models included. Compared to the flat weight $w_i = \left(\frac{x_i}{\sum_{j = 1}^{n} x_j}\right)$, the Softmax function gives more weight to the good models and almost neglects the bad ones due to the exponential term.

In Table \ref{Tab:errors} the performance of the three models are shown for the average and marginal CO\textsubscript{2} emission intensity (the response variable $\mathbf{y}$) when using the forecast horizon of $h = 6$ hours. Listed are RMSE's..

Model \textbf{M2} with the linear interaction terms is the best model for both the average and marginal emissions implying the importance of variable interactions. The marginal emissions have lower errors compared to the average emissions, suggesting that the marginal value is easier to predict - it is less influenced by highly uncertain (weather dependent) variables as already remarked.

The weighted average model $\text{M}_{\rm{WA}}$ is constructed by combining the models with the Softmax weights: the RMSE in the \emph{test} set becomes 38.56 and 9.63 $\left[\frac{\text{kgCO\textsubscript{2}-eq}}{\text{MWh}}\right]$ for the average and marginal emissions, respectively. This is only a slight improvement to the RMSE compared to \textbf{M2}, because of the large weight assigned to it.

\begin{table} \small%
\begin{tabular}{ l l | c  c  c | c}
 \textbf{Average}                &  & \textbf{M1} & \textbf{M2} & \textbf{M3} & \textbf{M\textsubscript{WA}}\\
\hline
\textbf{Validation } &    & 39.63 & 38.97 & 39.19 & \textbf{37.87}\\
 \textbf{Test}            &  & 41.13 & 39.54 & 40.37  & \textbf{38.45}\\
   \textbf{Weights}      & &\textbf{0.22} & \textbf{0.43} &  \textbf{0.35} & \\
\hline
                &  &  &  & & \\
 \textbf{Marginal}                &  &  &  &  &\\
 \hline
\textbf{Validation } &    & 11.06 & 8.77 & 10.03 & \textbf{8.57}\\
\textbf{Test}            & & 11.94 & 9.94 & 10.83  & \textbf{9.63}\\
  \textbf{Weights}       &   &\textbf{0.07} & \textbf{0.73} &  \textbf{0.2} & \\
\hline
\end{tabular} 
\caption{Root-Mean-Squared Error [RMSE] of the average and marginal emissions for models \textbf{M1}-\textbf{M1}, with a prediction horizon of 6 hours. The weights are calculated using the RMSE values from the validation sets. \textbf{M\textsubscript{WA}} is the resulting weighted average model. Units: $\left[\frac{\text{kgCO\textsubscript{2}-eq}}{\text{MWh}}\right]$. \label{Tab:errors}}
\end{table} 


\subsection{The compound model}
The pure $\text{M}_{WA}$ model of the previous section - i.e., without the residual correction of Sec. \ref{section:corrections} - was used for forecasts on each individual horizon $h = 1, 2,..., 24$ hours (denoted $\text{M}_{WA1...}$). Trials showed that ARIMA model performed on $h = 6$ hours corrects the residuals on all earlier horizons; however, individual $\rm{\textbf{M}}_{WA1,2}$ as well as $\rm{\textbf{M}}_{WA7-24}$ outperform the $\rm{\textbf{M}}_{ARIMA6}$ on their corresponding horizons for the average emissions. This is so both because individual models designed for specific horizons may perform better, and also ARIMA prediction converge towards the average as the prediction horizon increases, thus being less suitable for longer horizons. The models for different horizons with the corresponding RMSEs for the average and marginal emissions are summarized in Table \ref{tab:modelResults}. 



\begin{figure}[t]
\centering
  \includegraphics[width=\linewidth]{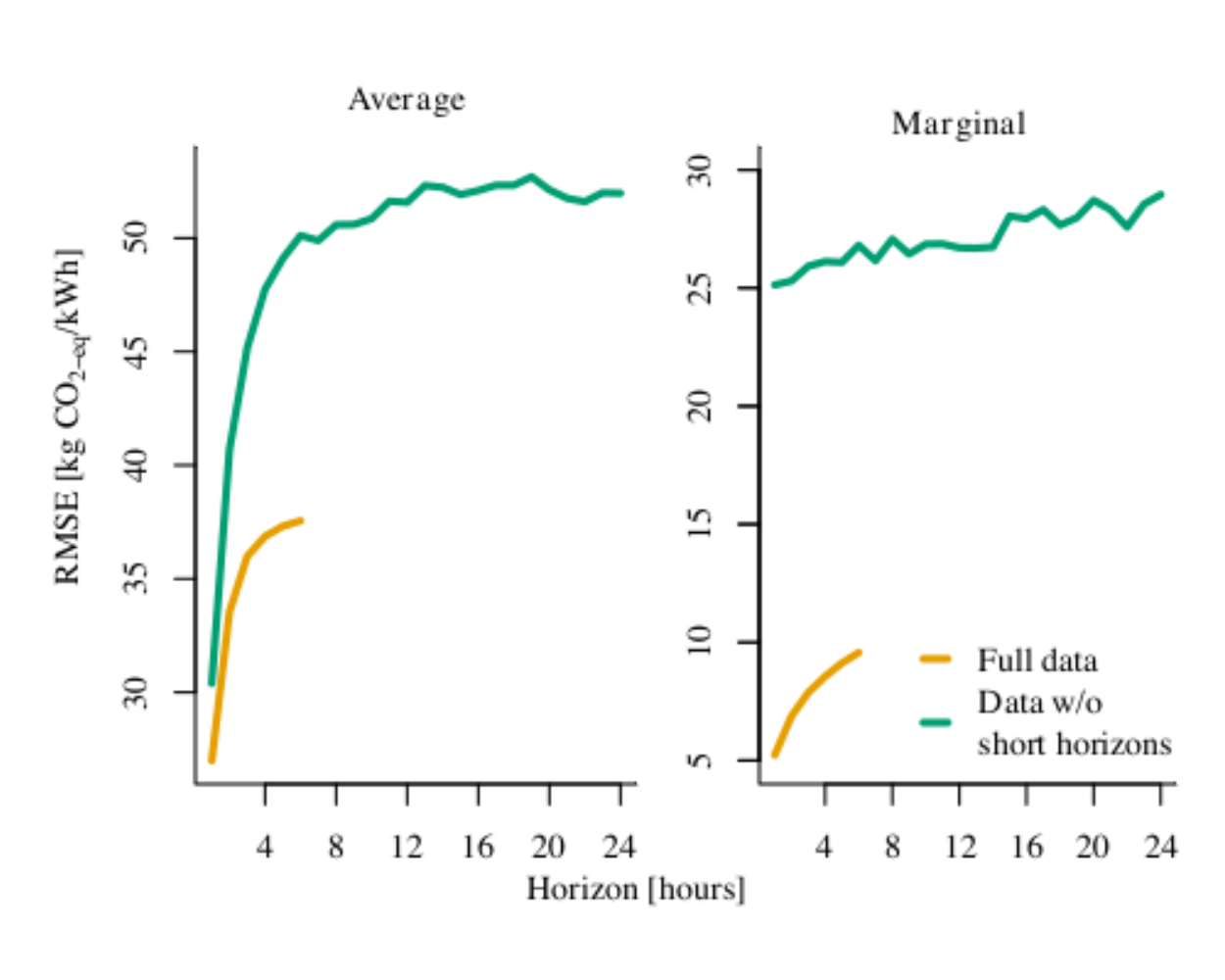}
  \caption{RMSE of CO\textsubscript{2} emission predictions vs prediction horizon. When short term forecasts are excluded from the data set (green line), the error increases (31\% on the 6th hour horizon for the average emissions). The error stays at around the same level for $h>6$ hours (49-52$\left[\frac{\text{kgCO\textsubscript{2}-eq}}{\text{MWh}}\right]$). In the marginal emissions, the short term forecasts are more important for the predictions and without them the error is high even on the short horizons.}
  \label{fig:hypthetical}
\end{figure}

\begin{table}[t]
    \small
    \centering
    \begin{tabular}{ m{1cm}  |m{0.6cm} m{0.8cm} m{0.9cm} | m{1cm} m{0.6cm} }
        &  \multicolumn{3}{c}{\textbf{Average}}  & \multicolumn{2}{c}{\textbf{Marginal}} \\[5pt]
                & \small{\textbf{M}\textsubscript{WA1,2}} & \shortstack[c]{\small{\textbf{M}\textsubscript{ARIMA6}}} &    \small{\textbf{M}\textsubscript{WA7-24}} & \shortstack[c]{\small{\textbf{M}\textsubscript{ARIMA6}}} & \textbf{M}\textsubscript{WA7-24} \\
                \hline
            \shortstack[c]{Horizons \\ $[\rm{h}]$}  & 1-2 & $\;$ 3-6 & 7-24 & 1-6  & 7-24 \\
            \shortstack[l]{RMSE \\ $\left[\frac{\text{kgCO\textsubscript{2}-eq}}{\text{MWh}}\right]$}  & 27.0-33.5 & 36.0-37.6 & 49.9-52.0 & 5.2-9.6 &  26.2-29.0
    \end{tabular}
    \caption{Performance of the final best models for particular horizons on test set. The average emission model is thus on $h=6$ improved from 38.45 to 37.6  $\left[\frac{\text{kgCO\textsubscript{2}-eq}}{\text{MWh}}\right]$ and the marginal is improved from 9.7 to 9.6 $\left[\frac{\text{kgCO\textsubscript{2}-eq}}{\text{MWh}}\right]$ (from Table \ref{Tab:errors}).
    }
    \label{tab:modelResults}
\end{table}

Note from the table that RMSE during 7-24 hours becomes almost stationary 
The reason is that on longer horizons, current information, say, on production data (available through short-term forecasts, \ref{Appendix:Variables}) affects the predictions and associated uncertainties much less than the available long-term information of e.g. weather. On shorter horizons, 0-6 hours, RMSE depends on short-term data and gradually increases in time until reaching the stationary value. The features can be seen in Figure \ref{fig:hypthetical}.

The final compound model $\rm{\textbf{M}}_{24}$ used for the 24-hour forecast of the emissions is the sum of the corresponding triplets of Table \ref{tab:modelResults}.
When the short-term forecast data is included in $\rm{\textbf{M}}_{24}$, the RMSE are naturally smaller than when it is excluded (yellow vs. green line, Figure \ref{fig:hypthetical}). The improvement is by 31\% (within the 0-6 hour horizon, for which short-term forecasts are available).

\begin{table}[t!]
    \centering
    \begin{tabular}{m{2cm} m{1.95cm}| p{0.75cm}  p{0.75cm} p{0.58cm} }
            $\mathbf{z}_i$ & $\mathbf{z}_j$ &  $\beta^{IA}_i$ & $\beta^{IA}_j$ & $\beta^{IA}_{i,j}$ \\
            \hline
        \textbf{Average} &  &   &  &  \\[6pt]
       \shortstack[l]{Production \\ in DK2.} & \shortstack[l]{Spline \\ (midday).} & 57.3 & 1.7 & -4.7 \\[13pt]
       \shortstack[l]{Production \\ in DK2.} & \shortstack[l]{Daily pattern.} & 57.3 & 3.3 & -2.3 \\[13pt]
       \shortstack[l]{Net export from \\DK1 to DE .} & \shortstack[l]{Spline \\ (midday).}  & -11.4 & 1.7 & 3.8\\[13pt]
       \shortstack[l]{Wind speed\\ in DK2.} & \shortstack[l]{Net export\\ from DK2 to \\ DE.} & -11.8 & -8.2 & 4.2 \\[13pt]
       \shortstack[l]{Offshore wind\\ in DK2.} & \shortstack[l]{Net export \\from SW3 to \\ SW4.}   &  -16.7 & -14.1 & 2.6\\[6pt]
       \hline
    \textbf{Marginal} &  &  & &  \\[6pt]
       \shortstack[l]{Net export to\\ DE from \\ DK1 - exp.} & \shortstack[l]{Wind speed \\ in DE.}   &  10.9 & -6.1 & -1.9 \\[6pt]
       \shortstack[l]{solar \\ radiation \\in DE.} & \shortstack[l]{Net export \\ from DK1 to \\NO2 - exp.}  & -1.7 & -1.3 & 1.8\\[13pt]
       \shortstack[l]{solar \\ radiation \\in DE.}  & \shortstack[l]{Net export \\from  DK1 to  \\SW3 - exp.} & -1.7 & -3.7  & 1.4 \\[13pt]
       \shortstack[l]{Net export from \\ DK1 to SW3.} & \shortstack[l]{Net export  \\ from SW3 to \\SW4 - exp.}   & -8.0 & 0.54 & -1.5\\[13pt]
       \shortstack[l]{Net export from \\ SW3 to SW4.} & \shortstack[l]{Demand \\ in SW4}  & 11.4 &  2.7 & -0.9\\[6pt]

    \end{tabular}
 \caption{Strongest interactions in Model 2 for the average and marginal emission. $z$, $i$ and $j$ correspond to the terms in Equation \ref{eq:interaction}. Note the original variables in \ref{Appendix:Variables} use "net import" rather than "net export". Switched here, to make it easier to relate to.}\label{Tab:marg_inter}
\end{table}

\renewcommand{\arraystretch}{1}

\begin{figure*}[t]
\centering
  \includegraphics[width=\linewidth]{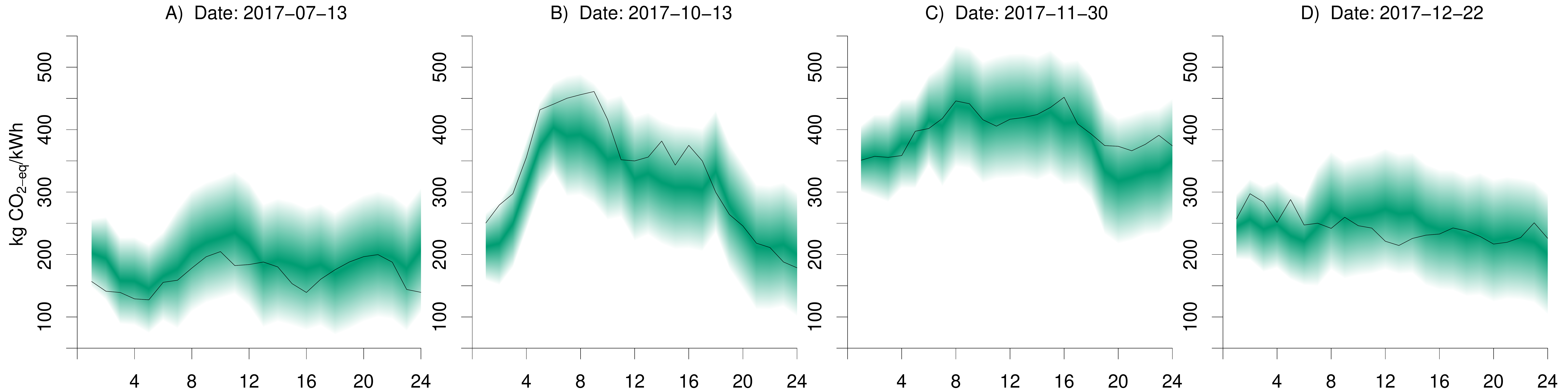}
  \caption{Average CO\textsubscript{2} emissions: The final forecast submitted at midnight with a 24 hour horizon for 8 different days. The real CO\textsubscript{2} emission intensity is the thick line, and the colored areas are bounding the 95\% confidence interval and the point prediction.}
  \label{fig:examples}
\end{figure*}

\begin{figure*}[t]
\centering
  \includegraphics[width=\linewidth]{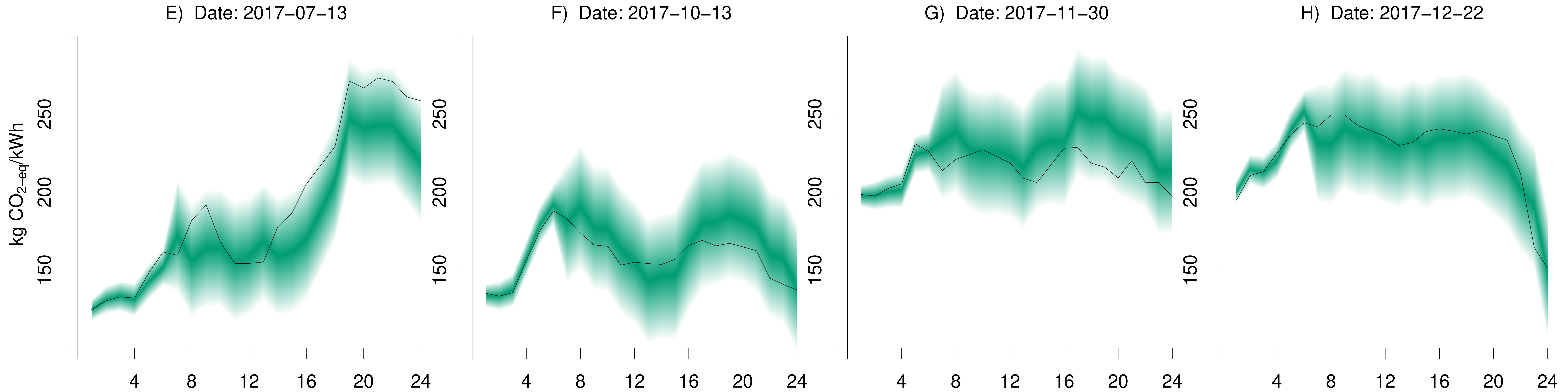}
  \caption{Marginal CO\textsubscript{2} emissions: The final forecast submitted at midnight with a 24 hour horizon for 8 different days. The real CO\textsubscript{2} emission intensity is the thick line, and the colored areas are bounding the 95\% confidence interval and the point prediction.}
  \label{fig:examples_mar}
\end{figure*}

\section{Selected results}\label{section:Results}

\subsection{The interaction coefficients $\mathbf{\beta}^{IA}$}
Of the three base models, \textbf{M2} that includes the interaction terms is the most accurate. The parameters and results of this model are discussed here for both the average and marginal emissions. Since the compound model puts the most weight on \textbf{M2}, the analysis applies to this model as well.

The five largest $\beta$ coefficients of \textbf{M2} are featured in Table \ref{Tab:marg_inter}, both for the average and the marginal emissions. $\beta^{IA}_i$, $\beta^{IA}_j$ and $\beta^{IA}_{i, j}$ from the table refer to the first, second and third coefficient of the interaction vector $\mathbf\beta^{IA}$ in Equation \ref{eq:interaction}. The minus sign indicates opposing trends. To understand the table properly please realize the following for interaction terms:
\begin{align}
    &\beta^{IA}_i \mathbf{z}_i + \beta^{IA}_j \mathbf{z}_j + \beta^{IA}_{i,j} \mathbf{z}_{i,j} \label{eq:beta1} \\
    = \; (&\beta^{IA}_i +\beta^{IA}_{i,j} \mathbf{z}_j)\cdot \mathbf{z}_i +  \beta^{IA}_j \mathbf{z}_j \label{eq:beta2}\\
    = \; (&\beta^{IA}_j +\beta^{IA}_{i,j} \mathbf{z}_i)\cdot \mathbf{z}_j +  \beta^{IA}_i \mathbf{z}_i \label{eq:beta3}
\end{align}

This means the final coefficient for e.g. $\mathbf{z}_i$ becomes $(\beta^{IA}_i +\beta^{IA}_{i,j} \mathbf{z}_j)$. The column for $\beta^{IA}_i$ is thus the coefficient for $\mathbf{z}_i$ if $\mathbf{z}_j$ is zero and vice versa. The column for $\beta^{IA}_{i,j}$ is the linear coefficient that explains the change in the final coefficient for both $\mathbf{z}_i$ and $\mathbf{z}_j$.

In the \textbf{average} emission model, the production in DK2 has the largest coefficient, 57.26: more production $\xrightarrow{}$ higher emissions, as also concluded in Figure~\ref{fig:pair}. It is interacting with the daily periodicities: the midday spline (row one in Table \ref{Tab:marg_inter}, see \ref{Appendix:tau}), and the daily pattern from Figure~\ref{fig:daily} (row five). The negative interaction coefficients $\beta^{IA}_{i,j}$, -4.7 and -2.3, via Equation \ref{eq:beta2}, mean the impact from the power production decays and reaches a minimum at around noon. This is because other factors start to influence the emissions more. 

The midday spline is again interacting with the net export from DK1 to DE (row three) that has a negative coefficient -11.4; here, $\beta^{IA}_{i,j}$ is positive, 3.8, and suggests the least impact during the midday, concluded from Equation \ref{eq:beta2}.

The wind speed in DK2 (row four) has a negative coefficient: the average emissions will decrease as more wind power enters the grid. The wind speed is positively interacting with net export from DK2 to DE (4.2) hence the final wind speed coefficient becomes larger, approaching zero, because the wind power is being exported rather than being used to decrease the emissions in DK2.

The offshore wind power production in DK2 (row five) also has a negative coefficient, -16.7, and is interacting with the net export from SW3 to SW4 - this is primarily a one way inter-connector explicitly exporting to SW4. The emissions in DK2 decrease as exports increase - this is expected because SW3 is in possession of all Sweden's nuclear power \cite{Kraftnat}, and that is exported further into DK2. $\beta^{IA}_{i,j}$ is positive and the final wind power coefficient will approach zero as the export increase, Equation \ref{eq:beta2}: the nuclear power takes part in the overall emissions in DK2 making the wind power contribution account for relatively less.

In the \textbf{marginal} emission model, note that none of the listed variables describe the grid in DK2. All are external influences from neighboring bidding zones and include net exports in all interactions. They are created because the net exports often only impact the emissions in one trading direction. Recall from Figure \ref{fig:pair_res}, only the exports from DK1 to DE has an impact on the emissions. In this model, the coefficient of that trading pair is positive (row six) fitting the exports. To force the coefficient closer to zero during import, it is negatively interacting with the wind speed in DE, $\beta^{IA}_{i,j} = -1.9$. DK1 will import if DE has wind power to offer and the final coefficient for the net export decrease.

The solar radiation in DE is both interacting with net exports from DK1 to NO2 and net exports DK1 to SW3 ($\beta^{IA}_{i,j}=1.8$ and  1.4 respectively). The two interactions explain the same phenomenon: the emissions decrease as DK1 exports to SW3 and NO2 because $\beta_j = -1.3$ is negative. However, when solar radiation in DE is high DK1 will again import from DE and the final coefficients for the net exports increase approaching zero because of the positive interaction coefficients.

The net export from DK1 to SW3 (row 9) is interacting with the net export from SW3 to SW4 (exponential term) too with $\beta^{IA}_{i,j} = -1.5$: export from SW3 to SW4 decreases the impact from the net export from DK1 to SW3. That is because the marginal emissions in SW3 increase as their total export increases.
The net export from SW3 to SW4 is the most significant variable and is negatively interacting with the demand in SW4 $\beta^{IA}_{i,j} = -0.9$: the final coefficient for the net import in SW4 from SW3 decreases as the demand increases, most likely because in this case power is consumed in SW4 rather than exported further into DK2. 

\subsection{CO\textsubscript{2} emissions: the forecast of $M_{24}$}
A few examples for the average and marginal emission forecasts by model $\rm{\textbf{M}}_{24}$ are shown in Figure~\ref{fig:examples} and Figure~\ref{fig:examples_mar} to illustrate the performance, where a 24-hour horizon forecast is released at midnight. Note, the dates in the plots for the average and marginal emissions are identical.

Comparing the plots (\textbf{average} emissions) to the daily pattern in Figure~\ref{fig:daily}, plot B and C fit best with the highest emissions during the day. Plot B peaks already in the morning slightly higher than the predictions and in plot C there is expected a lower decrease than observed in the evening. To a certain degree, daily patterns are often expected, so when the real observations differ too much, the accuracy decreases: in plot D, the emissions had a downward going trend all day, but it was expected to peak at around noon and then decrease. Plot A illustrates a day with irregularities too where the trend is captured to an adequate degree. 

The \textbf{marginal} emissions have a much slimmer confidence interval than the average emissions due to the higher accuracy. Plot F differs the least from the average daily pattern and the prediction shows this too. The predictions in Plot G had a low accuracy because of irregular and small spikes. Plot E and H are good examples for a control mechanism: in plot A, the emissions are expected to increase in the evening, so it is encouraged to schedule flexible demand as early as possible before the emissions increase. In plot D, the opposite is seen: here, the demand should be shifted to the late evening where the emissions decreased.

%

\section{Conclusion}
\label{Section:Conclusion}
From data collected and supplied by Tmrow IVS, new forecasting models for average and marginal CO\textsubscript{2} emissions in the European electricity grid are developed using linear regression and residual correction. A machine learning methodology to systematically select the important variables that best fit the desired variable is presented.

It is found that interactions between the explanatory variables are important: large coefficients are found for net imports, and time-dependence is least pronounced during midday or midnight. Interestingly, none of the most important variable related to the marginal emissions in DK2 were local (DK2) variables - all contributions came from neighboring bidding zones (DK1, DE, SW3 (indirect) and SW4). This suggests that the marginal generator is effectively supplied from the import, in agreement with \cite{Corradi} which found mainly import from SW4.

The study aimed to provide a tool that can help electricity consumers schedule their load to minimize CO\textsubscript{2} emissions. This was accomplished by forecasts of emissions 24 hours ahead, which provides a basis for decision making for load scheduling. The average and marginal emissions follow different patterns that can be exploited for different applications. The marginal CO\textsubscript{2} emissions are valid for small changes in demand and are therefore the signal to use when scheduling home appliances. The average emissions are useful for evaluating total electricity system emissions but should not be used as a control signal.

To evaluate the usefulness of the marginal emission forecast, testing on various flexible applications, e.g. heat pumps, electric cars, etc. should be conducted in the future. Results from this can indicate if there is a need for further model improvements. The marginal emission estimates used in this study cover most situations but there are still limitations as mentioned in the footnote \ref{foot:mar} (page 1). Further studies are needed to incorporate the weather dependent generators as marginals to fully understand the concept.

\section{Acknowledgement}
We are thankful for Tmrow IVS who has provided the data used in this study (including emission calculations for the bidding zone DK2). The work is supported through the project “Smart Cities Accelerator 2016-2020”  funded by the EU program Interreg Öresund-Kattegat-Skagerrak, the European Regional Development Fond and the CITIES project (DSF1305-00027B).

\bibliographystyle{elsarticle-num} 
\bibliography{elsarticle-template-num}

\begin{thebibliography}{10}
\expandafter\ifx\csname url\endcsname\relax
  \def\url#1{\texttt{#1}}\fi
\expandafter\ifx\csname urlprefix\endcsname\relax\def\urlprefix{URL }\fi
\expandafter\ifx\csname href\endcsname\relax
  \def\href#1#2{#2} \def\path#1{#1}\fi

\bibitem{IEA_emi}
C$\text{O}_2$ emissions statistics, \url{https://www.iea.org/}, [Online;
  accessed 6-Nov-2019] (2017).

\bibitem{IEA_sol}
Status of power system transformation 2019: Power system flexibility,
  \url{https://www.iea.org}, [Online; accessed 6-Nov-2019] (2019).

\bibitem{Zheng}
Z.~Liang, J.~Liang, C.~Wang, X.~Dong, X.~Miao, Short-term wind power combined
  forecasting based on error forecast correction, Applied Energy 119 (2016)
  215--226.

\bibitem{Wang}
J.~Wang, T.~Niu, H.~Lu, Z.~Guo, W.~Yang, P.~Du, An analysis-forecast system for
  uncertainty modeling of wind speed: A case study of large-scale wind farms,
  Applied Energy 211 (2018) 492--512.

\bibitem{Bacher}
P.~Bacher, H.~Madsen, H.~A. Nielsen, An analog ensemble for short-term
  probabilistic solar power forecast, Solar Energy 10 (2009) 1772--1783.

\bibitem{Delle}
S.~Alessandrini, L.~D. Monache, S.~Sperati, G.~Cervone, An analog ensemble for
  short-term probabilistic solar power forecast, Applied Energy 157 (2015)
  95--110.

\bibitem{Dogan}
D.~Keles, J.~Scelle, F.~Paraschiv, W.~Fichtner, Extended forecast methods for
  day-ahead electricity spot prices applying artificial neural networks,
  Applied Energy 162 (2016) 218--230.

\bibitem{Zhang}
Z.~Yang, L.~C. amd Li~Lian, Electricity price forecasting by a hybrid model,
  combining wavelet transform, arma and kernel-based extreme learning machine
  methods, Applied Energy 190 (2017) 291--305.

\bibitem{Wolfgang}
U.~Wagner, W.~Mauch, S.~von Roon, Das merit-order-dilemma der emissionen, Tech.
  rep., Forschungsstelle für Energiewirtschaft e.V (2002).

\bibitem{Anika}
A.~Regett, F.~Baing, J.~Conrad, Emission assessment of electricity: Mix vs.
  marginal power plant method, 15th International Conference on the European
  Energy Market (EEM) (2018).

\bibitem{Voors}
Voorspools, D'haeseleer, An evaluation method for calculating the emission
  responsibility of specific electric applications, Energy Policy 28 (2006)
  967--980.

\bibitem{Voors2}
Voorspools, D'haeseleer, The influence of the instantaneous fuel mix for
  electricity generation of the corresponding emissions, Energy 25 (2000)
  1119--1138.

\bibitem{Marnay}
Marnay, Fisher, Murtishaw, Phadke, Price, Sathaye, Estimating carbon dioxide
  emissions factors for the california electric power sector, Tech. rep.,
  Lawrence National Laboratory, Berkley USA (2002).

\bibitem{Beetle}
R.~Bettle, C.~Pout, E.~Hitchin, Interactions between electricity-saving
  measures and carbon emissions from power generation in england and wales,
  Energy Policy 34 (2006) 3434--3446.

\bibitem{Hawkes}
A.~Hawkes, Estimating marginal co2 emissions rates for national electricity
  systems, Energy Policy 38 (2010) 5977--5987.

\bibitem{Rekkas}
Rekkas, Uk marginal powerplant and emissions factors, Master's thesis, Imperial
  College London (2005).

\bibitem{Hadland}
Hadland, Marginal emissions factors for the united kingdom electricity system,
  Master's thesis, Imperial College London (2009).

\bibitem{CO2_signal_heating}
J.~Clauß, S.~Stinner, C.~Solli, K.~B. Lindberg, H.~Madsen, L.~Georges,
  Evaluation method for the hourly average co2-eq intensity of the electricity
  mix and its application to the demand response of residential heating,
  energies 12 (2019).

\bibitem{Corradi}
O.~Corradi.
\newblock
  https://medium.com/electricitymap/using-machine-learning-to-estimate-the-hourly-marginal-carbon-intensity-of-electricity-49eade43b421
  [online] (2018).

\bibitem{Bialek}
J.~Bialek, Tracing the flow of electricity, Vol. 143, IEEE, 1996.

\bibitem{Kirschen}
D.~Kirschen, R.~Allan, G.~Strbac, Contributions of individual generators to
  loads and flows, Transactions on Power System 12 (1997) 52--60.

\bibitem{Azim}
A.~Heydari, D.~A. Garcia, F.~Keynia, F.~Bisegna, L.~D. Santoli, Renewable
  energies generation and carbon dioxide emission forecasting in microgrids and
  national grids using grnn-gwo methodology, Applied Energy 159 (2019)
  154--159.

\bibitem{Kraftnat}
Elområde 1-4 (sn1-4) - statistik per månad 2017), \url{https://www.svk.se/},
  [Online; accessed 1-Nov-2019] (2017).

\bibitem{Springer}
R.~T. Trevor~Hastie, J.~Friedman, The Elements of Statistical Learning,
  Springer Series in Statistics, 2017.

\bibitem{Aike}
Porter, A.~S, West, Multiple regression: Testing and interpreting interactions,
  Journal of the Royal Statistical Society: Series D (The Statistician) 43
  (1994) 453.

\bibitem{Hu}
M.~Y. Hu, G.~Zhang, C.~X. Jiang, B.~E. Patuwo, A cross-validation analysis of
  neural network out-of-sample performance in exchange rate forecasting,
  Decision Sciences 30 (1999) 197--215.

\bibitem{Tibshirani}
R.~Tibshirani, Regression shrinkage and selection via the lasso: a
  retrospective, Journal of the Royal Statistical Society. Series B
  (Statistical Methodology) 73 (2011) 273--282.

\bibitem{TSA}
H.~Madsen, Time Series Analysis, Chapman \& Hall/CRC - Taylor \& Francis Group,
  2007.

\end{thebibliography}





\appendix
\newpage
\section{Explanatory variables} \label{Appendix:Variables}
Here, all explanatory variables are listed by data set for bidding zone DK2.
For each variable in the data sets, it is indicated whether the variable is used to create models for $h \leq 6$ hours or $h > 6$ hours.

\begin{table}[htbp]
\small
  \centering
    \begin{tabular}{lcc}
    \thead{Short Term Forecasts} & & \\
    & \multicolumn{1}{>{\small}l}{$h \leq 6$ hours} & \multicolumn{1}{>{\small}l}{$h > 6$ hours} \\
    dewpoint\ & X     &  \\
    precipitation\ & X     &  \\
    solar\ & X     &  \\
    temperature\ & X     &  \\
    price\ & X     &  \\
    production\ & X     &  \\
    consumption\ & X     &  \\
    wind\_speed\ & X     &  \\
    wind\_direction\_x\ & X     &  \\
    wind\_direction\_y\ & X     &  \\
    power\_net\_import\_DK-DK1\ & X     &  \\
    power\_net\_import\_DE\ & X     &  \\
    power\_net\_import\_SE-SE4\ & X     &  \\
    power\_net\_import\_SE\ & X     &  \\
\end{tabular}%
\label{tab:addlabel}%
\caption*{6 hours Forecasts (updated every sixth hour) provided by Tmrow IVS. Imports are originally coming from ENTSO-E who collect the data from individual Transmission System Operators (TSO's).}
\end{table}%

\begin{table}[htbp]
\small
  \centering
    \begin{tabular}{lcc}
    \thead{Weather Forecasts} & & \\
    & \multicolumn{1}{>{\small}l}{$h \leq 6$ hours} & \multicolumn{1}{>{\small}l}{$h > 6$ hours} \\
    dewpoint\_mean\_value &       & X \\
    precipitation\_mean\_value &       & X \\
    solar\_mean\_value &       & X \\
    temperature\_mean\_value &       & X \\
    wind\_mean\_value &       & X \\
\end{tabular}%
\label{tab:addlabel2}%
\caption*{52 hours Forecasts (updated every sixth hour) provided by Tomorrow. Originally created by GFS - Global Forecasting System}
\end{table}%

\begin{table}[H]
\small
  \centering
    \begin{tabular}{lcc}
    \thead{Market Data (Nordpool)} & & \\
    & \multicolumn{1}{>{\small}l}{$h \leq 6$ hours} &
    \multicolumn{1}{>{\small}l}{$h > 6$ hours} \\
    solar\_power & X     & X \\
    wind\_power\_offshore  & X     & X \\
    wind\_power\_onshore  & X     & X \\
    production &       & X \\
    consumption  &       & X \\
    spot\_price  &       & X \\
    \end{tabular}%
  \label{tab:addlabel3}%
  \caption*{Published at 6pm CET covering the whole next day.
Reported in at 12pm CET, and technically available at that time. }
\end{table}%


\begin{table}[H]
\small
  \centering
    \begin{tabular}{lcr}
    \thead{Real Time Data} & & \\
          & \multicolumn{1}{>{\small}l}{$h \leq 6$ hours} & \multicolumn{1}{>{\small}l}{$h > 6$ hours} \\
    carbon\_intensity\ & X     & \multicolumn{1}{c}{X} \\
    carbon\_intensity\_production\ & X     & \multicolumn{1}{c}{X} \\
    carbon\_intensity\_import\ & X     & \multicolumn{1}{c}{X} \\
    carbon\_rate\ & X     & \multicolumn{1}{c}{X} \\
    total\_production\ & X     & \multicolumn{1}{c}{X} \\
    total\_storage\ & X     & \multicolumn{1}{c}{X} \\
    total\_discharge\ & X     & \multicolumn{1}{c}{X} \\
    total\_import\ & X     & \multicolumn{1}{c}{X} \\
    total\_export\ & X     & \multicolumn{1}{c}{X} \\
    total\_consumption\ & X     & \multicolumn{1}{c}{X} \\
    power\_origin\_\%\_fossil\ & X     & \multicolumn{1}{c}{X} \\
    power\_origin\_\%\_renewable\ & X     & \multicolumn{1}{c}{X} \\
    power\_production\_biomass\ & X     & \multicolumn{1}{c}{X} \\
    power\_production\_coal\ & X     & \multicolumn{1}{c}{X} \\
    power\_production\_gas\ & X     & \multicolumn{1}{c}{X} \\
    power\_production\_hydro\ & X     & \multicolumn{1}{c}{X} \\
    power\_production\_nuclear\ & X     & \multicolumn{1}{c}{X} \\
    power\_production\_oil\ & X     & \multicolumn{1}{c}{X} \\
    power\_production\_solar\ & X     & \multicolumn{1}{c}{X} \\
    power\_production\_wind\ & X     & \multicolumn{1}{c}{X} \\
    power\_production\_geo\ & X     & \multicolumn{1}{c}{X} \\
    power\_production\_unknown\ & X     & \multicolumn{1}{c}{X} \\
    power\_origin\_\%\_biomass\ & X     & \multicolumn{1}{c}{X} \\
    power\_origin\_\%\_coal\ & X     & \multicolumn{1}{c}{X} \\
    power\_origin\_\%\_gas\ & X     & \multicolumn{1}{c}{X} \\
    power\_origin\_\%\_hydro\ & X     & \multicolumn{1}{c}{X} \\
    power\_origin\_\%\_nuclear\ & X     & \multicolumn{1}{c}{X} \\
    power\_origin\_\%\_oil\ & X     & \multicolumn{1}{c}{X} \\
    power\_origin\_\%\_solar\ & X     & \multicolumn{1}{c}{X} \\
    power\_origin\_\%\_wind\ & X     & \multicolumn{1}{c}{X} \\
    power\_origin\_\%\_geo\ & X     & \multicolumn{1}{c}{X} \\
    power\_origin\_\%\_unknown\ & X     & \multicolumn{1}{c}{X} \\
    power\_origin\_\%\_hydro\ & X     & \multicolumn{1}{c}{X} \\
    carbon\_origin\_\%\_biomass\ & X     & \multicolumn{1}{c}{X} \\
    carbon\_origin\_\%\_coal\ & X     & \multicolumn{1}{c}{X} \\
    carbon\_origin\_\%\_gas\ & X     & \multicolumn{1}{c}{X} \\
    carbon\_origin\_\%\_hydro\ & X     & \multicolumn{1}{c}{X} \\
    carbon\_origin\_\%\_nuclear\ & X     & \multicolumn{1}{c}{X} \\
    carbon\_origin\_\%\_oil\ & X     & \multicolumn{1}{c}{X} \\
    carbon\_origin\_\%\_solar\ & X     & \multicolumn{1}{c}{X} \\
    carbon\_origin\_\%\_wind\ & X     & \multicolumn{1}{c}{X} \\
    carbon\_origin\_\%\_geo\ & X     & \multicolumn{1}{c}{X} \\
    carbon\_origin\_\%\_unknown\ & X     & \multicolumn{1}{c}{X} \\
    carbon\_origin\_\%\_hydro\ & X     & \multicolumn{1}{c}{X} \\
    power\_net\_discharge\_hydro\ & X     & \multicolumn{1}{c}{X} \\
    power\_net\_import\_DK-DK1\ & X     & \multicolumn{1}{c}{X} \\
    power\_net\_import\_DE\ & X     & \multicolumn{1}{c}{X} \\
    power\_net\_import\_SE-SE4\ & X     & \multicolumn{1}{c}{X} \\
    power\_net\_import\_SE\ & X     & \multicolumn{1}{c}{X} \\
    \end{tabular}%
  \label{tab:addlabel4}%
  \caption*{Provided by Tomorrow and originally created by GFS.}
\end{table}%

\newpage
\newpage

\section{Periodic time variables} \label{Appendix:tau}
\begin{figure}[t]
\centering
  \includegraphics[width=\linewidth]{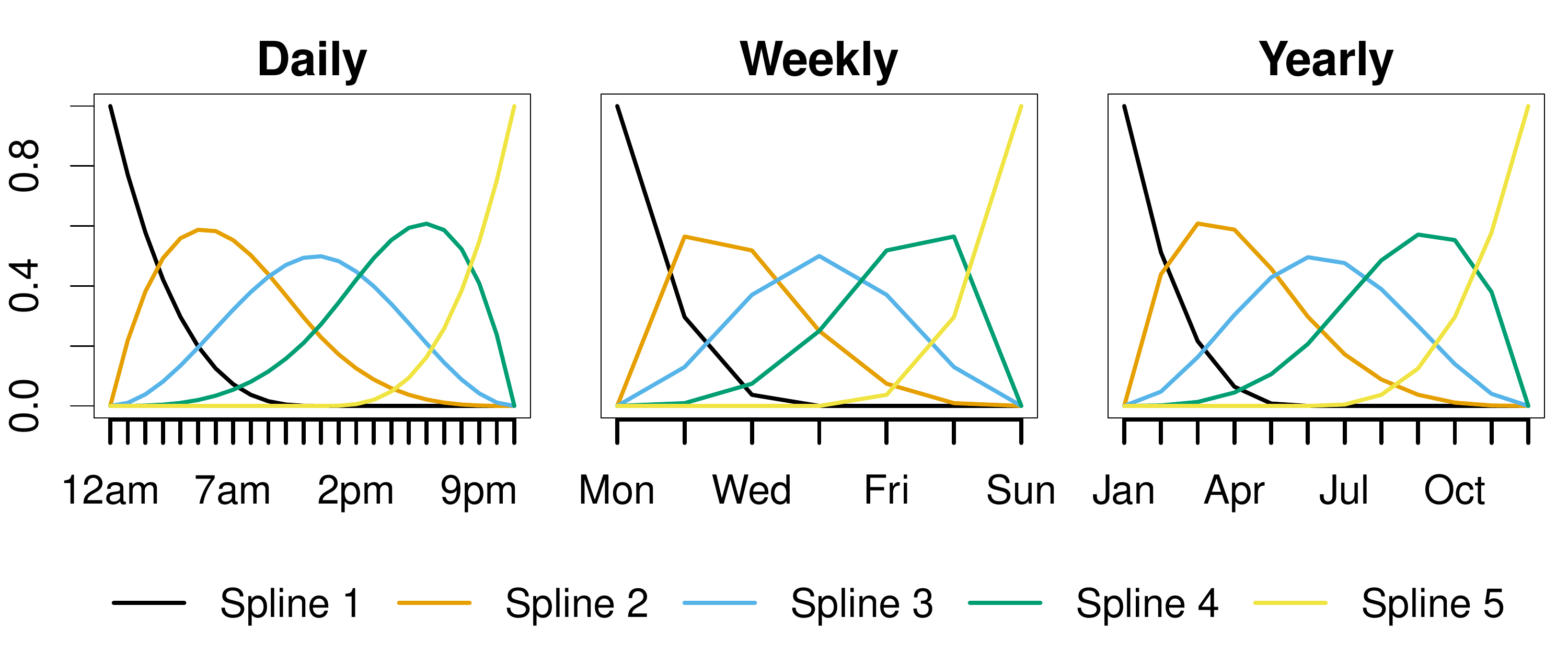}
  \caption{Periodic splines - Daily, weekly and yearly. Spine 3 in Daily is referred to as the midday spline. The outer splines (1 and 5) are not included in the model because they can be misleading.}
  \label{fig:periodic_splines}
\end{figure}

Time variable matrix $\boldsymbol{\tau}$ is defined as:
\begin{align} 
\boldsymbol{\tau}_{\text{hour}} &= t \text{ mod } 24 \nonumber \\
    \boldsymbol{\tau}_{\text{w}} &= \text{weekday}(t) \nonumber \\
    \boldsymbol{\tau}_{\text{m}} &= \text{month}(t) \nonumber \\
        \boldsymbol{\tau}_{\text{sin,hour}} &= \text{sin}(\boldsymbol{\tau}_{\text{hour}}) \nonumber \\
    \boldsymbol{\tau}_{\text{sin,w}} &= \text{sin}(\boldsymbol{\tau}_{\text{w}}) \nonumber \\
    \boldsymbol{\tau}_{\text{sin,m}} &= \text{sin}(\boldsymbol{\tau}_{\text{m}}) \nonumber \\
    \boldsymbol{\tau}_{\mathit{bs},\text{hour},t} &= \begin{bmatrix}
 \mathit{bs}_0(\boldsymbol{\tau}_{\text{hour},t}) & \mathit{bs}_1(\boldsymbol{\tau}_{\text{hour},t}) & ... &
 \mathit{bs}_{n,\text{df} - 1}(\boldsymbol{\tau}_{\text{hour},t})
 \end{bmatrix} \nonumber \\
     \boldsymbol{\tau}_{\mathit{bs}, \text{w},t} &= \begin{bmatrix}
 \mathit{bs}_0(\boldsymbol{\tau}_{\text{w},t}) & \mathit{bs}_1(\boldsymbol{\tau}_{\text{w},t}) & ... &
 \mathit{bs}_{n,\text{df} - 1}(\boldsymbol{\tau}_{\text{w},t})
 \end{bmatrix} \nonumber \\
 \boldsymbol{\tau}_{\mathit{bs}, \text{m},t} &= \begin{bmatrix}
 \mathit{bs}_0(\boldsymbol{\tau}_{\text{m},t}) & \mathit{bs}_1(\boldsymbol{\tau}_{\text{m},t}) & ... &
 \mathit{bs}_{n,\text{df} - 1}(\boldsymbol{\tau}_{\text{m},t})
 \end{bmatrix}, \nonumber
\end{align}
where hour, w and m denote the hour, weekday and month of the datetime $t$,  respectively.
$\boldsymbol{\tau}_{\mathit{bs}.\text{hour}}$, $\boldsymbol{\tau}_{\mathit{bs},\text{w}}$ and $\boldsymbol{\tau}_{\mathit{bs},\text{m}}$ each represent five columns corresponding to their underlying splines. $n_{df}$ is the number of splines which is set to 5 in this case. The periodic splines are illustrated in Figure \ref{fig:periodic_splines}.

\newpage
\section{Forward feature selection}  \label{Appendix:FS}
The forward selection algorithm selects the best variables for a model and requires a good cross validation strategy to avoid overfitting.
\begin{algorithm}[htbp]
\begin{enumerate}
    \item 1) Find the variable that best describes the response variable. This can be done with any best fit criteria (BIC, AIC or RMSE). This study relies on the RMSE value calculated on the validation sets of Sec. \ref{section:CV}. Call this Model\textsubscript{best}.
    \item 2) Add a new variable $\mathbf{x}_i$
    \begin{align}
        \mathbf{y}_{t+h} = \beta_0 + \beta_1 \mathbf{x}_{\mathit{best}} + \beta_2 \mathbf{x}_i. 
    \end{align}
    Call this Model\textsubscript{new}.
    \item 3) Evaluate the model. If Model\textsubscript{new} is better than  Model\textsubscript{best}, keep the newly added variable and update:  $\text{Model}_{\text{best}} = \text{Model}_{\text{new}}$.
    \item 4) Repeat step 2 and 3 until all variables have been tested.
\end{enumerate}
 \caption{Forward feature selection \label{alg:FS}}
\end{algorithm} 

\end{document}